\documentclass[aps,pre,twocolumn,groupedaddress,superscriptaddress]{revtex4-1}
\usepackage{epsfig,amssymb,amsmath,graphicx,subfigure,hyperref}
\usepackage{tabularx}
\usepackage{float}
\usepackage{setspace}

\usepackage{array} 
\newcommand{\PreserveBackslash}[1]{\let\temp=\\#1\let\\=\temp} \newcolumntype{C}[1]{>{\PreserveBackslash\centering}p{#1}} \newcolumntype{R}[1]{>{\PreserveBackslash\raggedleft}p{#1}} \newcolumntype{L}[1]{>{\PreserveBackslash\raggedright}p{#1}}

\begin{document}

\title{Effects of scars on crystalline shell stability under external pressure  }

\author{Duanduan Wan}
\affiliation{Department of Physics, Syracuse University, Syracuse NY 13244, USA}

\author{Mark J.~Bowick}
\email[Correspondence should be sent to:]{bowick@phy.syr.edu}
\affiliation{Department of Physics, Syracuse University, Syracuse NY 13244, USA}
\affiliation{Syracuse Biomaterials Institute, Syracuse University, Syracuse NY 13244, USA}

\author{Rastko Sknepnek}
\affiliation{Division of Physics and Division of Computational Biology, University of Dundee, Dundee DD1 4HN, UK}

\date{\today}

\begin{abstract}
We study how the stability of spherical crystalline shells under external pressure is influenced by the defect structure. In particular, we compare stability
for shells with a minimal set of topologically-required defects to shells with extended defect arrays (grain boundary ``scars"
with non-vanishing net disclination charge). We perform Monte Carlo simulations to compare how shells with and without scars deform quasi-statically
under external hydrostatic pressure. We find that the critical pressure at which shells collapse is lowered for scarred configurations that break icosahedral symmetry
and raised for scars that preserve icosahedral symmetry. The particular shapes which arise from breaking of an initial icosahedrally-symmetric shell depend on the F{\" o}ppl-von K{\'a}rm{\'a}n number.   
\end{abstract}


\maketitle

\section{Introduction}

Thin elastic shells with spherical topology are ubiquitous in nature. Examples span a vast range of length scales: viral capsids \cite{fields2013virology} and nanocages \cite{huang1999nanocages}
at the nano-scale, pollen grains \cite{katifori2010foldable} at the micro-scale, ping-pong balls \cite{couturier2013folding} and stadium domes 
in the centimeter to decameter range and the Earth's crust on the scale of the entire planet. Despite the breadth of scales, all such shells may be considered thin, as measured by their thickness $h$ in terms of their linear size $R$ (i.e.,~$h/R\ll1$). 
As for thin plates, the mechanics of shells can be described in a continuum approach that expands the elastic energy in powers of the thickness \cite{audoly2010elasticity}. The resulting theory is non-linear even in the limit of small strains, as bending deformations ($O(h^3)$) \cite{van2009wt} play a significant role. In contrast to plates, however, any deformation of a shell couples stretching and bending \cite{LandauVol7,pogorelov1988bendings}, with remarkable consequences for the effect of thermal fluctuations on the material properties of the shell \cite{paulose2012fluctuating}. This is a consequence of the so-called \emph{geometric rigidity} of the sphere \cite{pogorelov1988bendings} - any deformation of the sphere is accompanied by a change of the local Gaussian curvature. The Gaussian curvature and the metric of a surface are intimately related (as recognized 
over two centuries ago by Gauss \cite{do1976differential}), and any change of Gaussian curvature will lead to changes in the metric, resulting in stretching or compression.

The situation is particularly striking if the shell is small enough that its crystalline structure plays a role. It is well known that 
perfect crystalline order is incompatible with spherical topology. In other words, it is not possible to cover a two-sphere with identical particles 
such that each one has exactly six equidistant nearest neighbors. Instead one needs to introduce topological defects - special points that have
coordination number different than six. More precisely, topology requires that $\sum_{i}\left(6-c_{i}\right)=6\chi=12 $ for any triangulation of a two-sphere,
where $c_{i}$ is the coordination number of vertex $i$ and $\chi=2$ is the Euler characteristic of the two-sphere. The quantity $q_{i}=6-c_{i}$ is called
the disclination charge and measures the departure from the perfect triangular order of the plane. Provided we restrict ourselves to the energetically 
preferred charge $q=\pm1$ disclinations, the topological constraint becomes $N_{+1}-N_{-1}=12$ and we see that a spherical crystal must have at least twelve $+1$ disclinations. These disclinations may also be viewed as singular concentrations of Gaussian curvature and in essence screen it \cite{sachdev1984crystalline}, thus lowering the elastic stress. With the defects being inherently discrete this screening cannot be perfect and the crystal will retain a certain level of residual stress even in the ground state. This has profound consequences for the mechanical behavior of the shell.

Limited to the minimal set of defects, the twelve disclinations repel each other, much like electric charges in two dimensions, and arrange themselves to minimize the
total energy. For an icosadeltahedral triangulation, the defects are located at the twelve vertices of an inscribed icosahedron and all possible crystal lattices
can be constructed following the prescription of Caspar and Klug \cite{caspar1962physical}. In this approach each lattice is labelled by a pair of
integers, $(p,q)$ that together form the $T$-number, $T=p^2+pq+q^2$, with the total number                of lattice sites being given by $N=10T+2$. For a sufficiently large number of particles, however, and provided the defect core energy is not very high, the ground state of a spherical crystal will have finite-length grain boundary \emph{scars} 
of tightly bound $5-7$ pairs radiating from the original $+1$ disclinations \cite{bowick2000interacting,bowick2002crystalline}. This phenomenon was also observed earlier \cite{perez1997influence} in a study of the Thomson problem \cite{thomson1904} of determining the ground state of classical electrons 
interacting with a repulsive Coulomb potential on the surface of a two-sphere. Scars have been observed experimentally in colloidal suspensions on spherical 
droplets - colloidosomes \cite{bausch2003grain, einert2005grain} - and in bubble rafts on a paraboloidal surface \cite{bowick2008bubble}.

If allowed to deform the shell can relieve some of the residual stress caused by the disclination defects by buckling. This was first discussed by Seung and Nelson \cite{seung1988defects} in 
the context of a single planar disclination. They showed that for a sufficiently large disclination radius $R$ it is energetically favorable to buckle into a cone with the apex at the disclination. The system reduces its stretching energy at the expense of gaining some bending energy. The transition occurs as the ratio of the energy scales that control the relative strength of bending vs.~stretching, the so-called F{\" o}ppl-von K\'{a}rm\'{a}n number $\gamma =  YR^2/\kappa\propto \left(\frac{R}{h}\right)^2$, exceeds a critical value of $\approx154$ ($Y\propto h$ is the two-dimensional (2D)
Young's modulus and $\kappa\propto h^3$ is the bending rigidity). On a sphere, the same mechanism drives a simultaneous buckling of all 12 disclinations leading to a transition from a sphere 
to an icosahedron \cite{lidmar2003virus}. The transition is rounded compared to the flat case, and shifted to slightly higher values of $\gamma$. Buckling into an icosahedron is quite robust and is not qualitatively affected by imposing volume constants \cite{siber2009stability} or by the presence of scars \cite{funkhouser2013topological}. 

The stability of thin crystalline shells under external pressure is at present only partly understood. Physical examples include suspensions of viruses in solution \cite{siber2009stability} and drug-delivery microcapsules in a flow \cite{varde2004microspheres}. The elastic theory of shells predicts the mechanical response of an ideal spherical shell
under external pressure \cite{van2009wt}. On microscopic scales crystalline order and defects may affect the stability of crystalline shells under external pressure. 

The paper is organized as follows. In Section \ref{sec:model} we present a discrete model for a shell under pressure. In Section \ref{sec:res} we analyze this model using Monte Carlo simulations and discuss how the critical pressure at which the shell collapses is related to the symmetry of the scars. In Section \ref{sec:discussion} we use spherical harmonic expansion to characterize symmetries of collapsing shells. Finally, in Section \ref{sec:summary} we summarize the main results and reflect on the possible extensions of this work.

\section{Model}
\label{sec:model}

In thin plate elasticity theory, a deformation is represented by an in-plane displacement vector field $\bold{u}\left(\bold{r} \right)=\left(u_{1}, u_{2} \right)$ and an out-of-plane displacement (``deflection") field $f\left(\bold{r} \right)$, which map the point $\left(x_{1}, x_{2}, 0\right)$ in the reference state to $\left(x_{1}+u_{1}, x_{2}+u_{2}, f\right)$. The elastic energy of an isotropic thin plate is the sum of a stretching and a bending energy \cite{LandauVol7,audoly2010elasticity}, $F_{el} = F_{s} + F_{b}$. In the regime of linear response, i.e.,~for small strains, the stretching energy is
\begin{equation}
F_{s} =\frac{1}{2} \int dS \left( 2 \mu u_{ij}^{2} + \lambda u_{kk}^{2} \right) \label{Fs}, 
\end{equation}
where 
\begin{equation}
u_{ij}=\frac{1}{2} \left(\partial_{i}u_{j} + \partial_{j}u_{i} + \partial_{i}u_{k}\partial_{j}u_{k} +\partial_{i}f\partial_{j}f \right)
\end{equation}
is the exact form of the strain tensor (indices run over 1 and 2) and the displacement fields are evaluated along the center surface. For small displacement gradients, the terms quadratic in $u_{k}$ could be neglected but the term quadratic in $f$ needs to be retained as there is no term of lower order in $f$. $\mu$ and $\lambda$ are the 2D Lam{\' e} coefficients
and the integral is over the area of the reference state. The 2D Young's modulus $Y=Eh$ ($E$ being the full three-dimensional Young's modulus) and Poisson ratio
$\nu$ can be expressed in terms of the 2D Lam\'{e} coefficients as \cite{seung1988defects}
\begin{equation}
Y = \frac{4 \mu \left( \mu + \lambda \right)}{2 \mu + \lambda} ~,  \qquad \nu = \frac{\lambda}{2 \mu + \lambda}   ~.
\end{equation} 
The bending energy is
\begin{equation}
F_{b} = \frac{1}{2}\kappa  \int dS \left( \left(\nabla^{2} f \right)^{2} - 2\left(1-\nu \right) \det\left(\partial_{i}\partial_{j} f \right) \right)
\label{Fb}, 
\end{equation}
where $\kappa=Eh^{3}/12\left(1-\nu^{2}\right)$ \cite{audoly2010elasticity} is the bending rigidity. In terms of $f$, twice the mean curvature $H$ and Gaussian curvature $K$ ($H=1/R_{1}+1/R_{2}$,  $K=1/R_{1}R_{2}$, with $R_{1}$ and $R_{2}$ being the principal radii of curvature, respectively) can be written as \cite{dubrovin1984geometry}  
\begin{equation}
H=\nabla \cdot \left( \frac{\nabla f}{\sqrt{1+|\nabla f|^{2}}} \right),  ~~~
K=\frac{\det \left(\partial_{i}\partial_{j} f\right)}{\left(1+|\nabla f|^{2} \right)^{2}} ~.  
\end{equation}
For small $|\nabla f|$,
\begin{equation}
H \approx \nabla^{2}f, ~~~~ K \approx \det \left(\partial_{i}\partial_{j} f\right),
\end{equation}
and the bending energy $F_{b}$ can be rewritten in the form of the Helfrich bending energy as 
\begin{equation}
F_{b} \approx \int dS \left( \frac{1}{2}\kappa  H^{2} + \kappa_{G} K \right)
\label{Fb_Helfrich},
\end{equation}
where $\kappa_{G}= -Eh^{3}/12\left(1+ \nu \right)$ is the Gaussian rigidity.
  
Within a discretized model where the flat plate is represented as a perfect triangular lattice, upon deformation, the stretching energy is given as \cite{seung1988defects,lidmar2003virus,schmidt2012universal} 
\begin{equation}
F_{s} = \frac{\varepsilon}{2}\sum_{\left< ij \right>}\left(\left|\mathbf{r}_{i}-\mathbf{r}_{j} \right| -a \right)^{2}
\label{Fs_dis},
\end{equation}
and the bending energy is given as
\begin{equation}
F_{b} = \frac{\tilde{\kappa}}{2}\sum_{\left< I J \right>}\left(\mathbf{\hat{n}}_{I}-\mathbf{\hat{n}}_{J}\right)^{2}
\label{Fb_dis},
\end{equation} 
where $\left<ij\right>$ denote pairs of nearest neighbor vertices, with positions $\mathbf{r}_{i}$ in the embedding three-dimensional space. $a$ is the equilibrium distance (the spacing in the flat state) and $\varepsilon$ is the elastic spring constant. $\tilde{\kappa}$ is the discrete bending modulus and $\left<IJ\right>$ denote pairs of nearest neighbor triangular plaquette, with unit normals $\mathbf{\hat{n}}_{I}$. In the continumm limit of a sufficiently large number of triangular plaquettes, it has been shown that in infinitesimal elasticity regime $Y= 2\varepsilon/\sqrt{3}$, $\nu = 1/3$  \cite{seung1988defects,schmidt2012universal} and for isometric immersions, Eq.~(\ref{Fb_dis}) becomes  \cite{schmidt2012universal}  
\begin{equation}
F_{b} = \frac{\tilde{\kappa}}{4\sqrt{3}}\int dS \left(3 H^2 -8K \right),
\label{Fb_dis_1}
\end{equation} 
which has the form of Eq.~(\ref{Fb_Helfrich}), with $\kappa = \sqrt{3}\tilde{\kappa}/2$ and $\kappa_{G} = -4\kappa /3$. Eq.~(\ref{Fb_dis_1}) provides a good approximation to the energy in bending-dominated deformation regime (an almost isometric immersion, a perfect isometric immersion implies pure bending and $K\equiv 0$) and general surfaces can be triangulated with an almost isometric immersion by a partition of the surface into regions of area on a mesoscopic scale and allowing for defects in the reference lattice along the boundary of these regions \cite{schmidt2012universal}.

In the numerical simulation we consider a discrete triangulated shell with the elastic energy being a sum of the stretching and bending energies given in Eqs.~(\ref{Fs_dis}) and (\ref{Fb_dis}). Shells without scars are constructed using the Caspar-Klug procedure (i.e., in terms of $(p,q)$ pairs), with the equilibrium distance $a$ being chosen as the mean value of all edge lengths over the entire triangulation (due to spherical topology not all edges can have equal length) and radii rescaled such that $a=1$. Shells with scars have the same total number of vertices and initial radii as shells without scars, with the connectivity matrix taken from the Thomson problem database \cite{thomson_applet}. 
To eliminate possible effects of thermal fluctuations, i.e., in the zero temperature limit, we set $k_{B}T = 10^{-11}$ and $\tilde{\kappa}=1$ (in units of energy) in all simulations. The low energy configurations are found using the Monte Carlo method for different values of $\varepsilon$ (in unit of $\text{energy}/a^2$). During the simulation the connectivity of the triangular mesh is held fixed as the shell's shape changes. In other words, we assume that the defects are frozen, consistent with the much slower defect dynamics
compared to shape relaxation \cite{yong2013elastic}.

\section{Numerical results}
\label{sec:res}
With the triangulation for shells without scars built following the $(p,q)$ structure and the connectivity matrix for shells with scar defects being that of the lowest energy configuration found for the Thomson problem,
we start by noting that, in the absence of external pressure, the equilibrated scarred shells have lower elastic energy compared to the equilibrated non-scarred shells for small values of $\varepsilon$. 
As $\varepsilon$ increases, shells without scars may release some stress by buckling at the twelve disclinations \cite{lidmar2003virus} and find configurations with lower elastic energy than those formed by shells with scars.

As we do not focus on the global energy minimum state of shells in the absence of external pressure, as studied in Ref.~\cite{funkhouser2013topological} or its evolution as a function of external pressure, which would require simultaneous relaxation of both the topography and connectivity (i.e., allow for defect motion), we only study
how the symmetry of the scar distribution in the initial configuration affects shell's tendency to succumb to pressure. We assume that the crystalline lattice on shells without scars always have 
icosahedral symmetry while shells with scars may break this symmetry. We therefore compare how shells without and with scars deform if the pressure is applied quasi-statically. We present results of shells with $N_{1}=642$ ($(p,q)=(8,0)$) and $N_{2}=1212$ ($(p,q)=(11,0)$) as generic
examples of cases where scars, respectively, break and preserve the icosahedral symmetry.

\begin{figure}
\centering 
\includegraphics[width=2.5in]{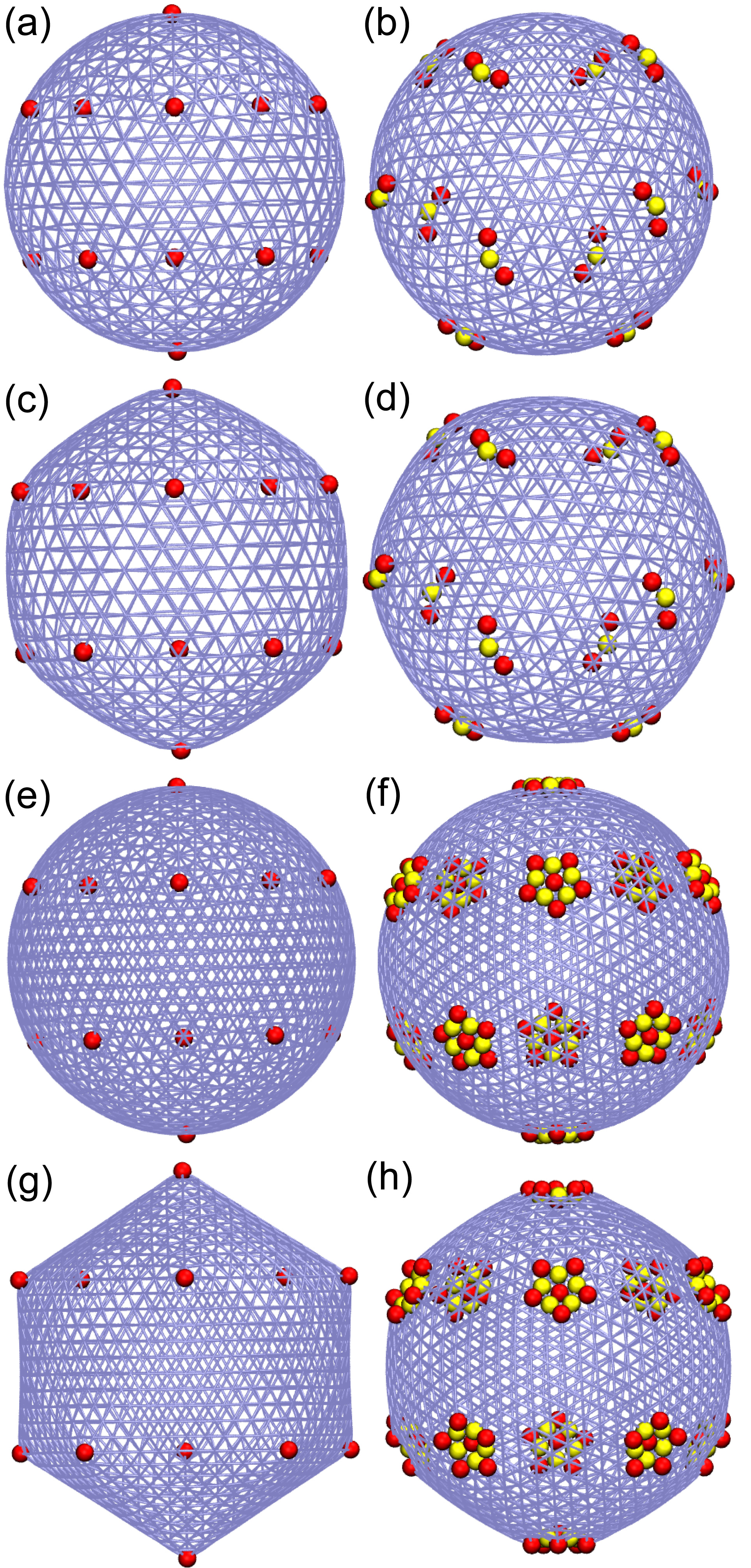}

\caption{(Color online) (a)-(d) Shells without scars have a $\left(p,q\right)=\left(8, 0\right)$ lattice structure (left) and shells with scars have twelve 
$5-7-5$ scars of $C_{2}$ symmetry (axis along the $z$ direction) (right), with $\varepsilon=1$ ($\gamma \approx 59$) (up) and $\varepsilon=8$ ($\gamma \approx 474$) (down). (e)-(h) Shells without scars have a $\left(p,q\right)=\left(11,0\right)$ structure (left) and shells with scars have twelve star-like scars of $I$ symmetry 
(right), with $\varepsilon=1$ ($\gamma \approx 112$) (up) and $\varepsilon=64$ ($\gamma \approx 7149$) (down). The total number of vertices is $N_{1}=642$ and $N_{2}=1212$, respectively. 
Five-fold disclinations are shown in red (dark) and seven-fold in yellow (bright). Snapshots were generated using the Visual Molecular Dynamics (VMD) package \cite{HUMP96} and rendered using the Tachyon ray tracer \cite{STON1998}.} 
\label{shells}
\end{figure}

Fig.~\ref{shells} shows some examples of equilibrated shells in the absence of external pressure. In the $N_{1}=642$ case, the scar-free lattice has an $\left(8,0\right)$ structure, while the shells with scars have twelve five-seven-five scars
of $C_{2}$ symmetry due to their mutual orientation. In the $N_{2}=1212$ case, shells without scars have an $\left(11,0\right)$ structure while shells with scars
have twelve identical star-like scars which preserve the icosahedral symmetry. For shells without scars the mean ``asphericity" \cite{lidmar2003virus} (i.e., the deviation from a perfect spherical shape) departs significantly from zero when $\gamma\gtrsim10^2$.

\begin{figure*}
\centering 
\includegraphics[width=3in]{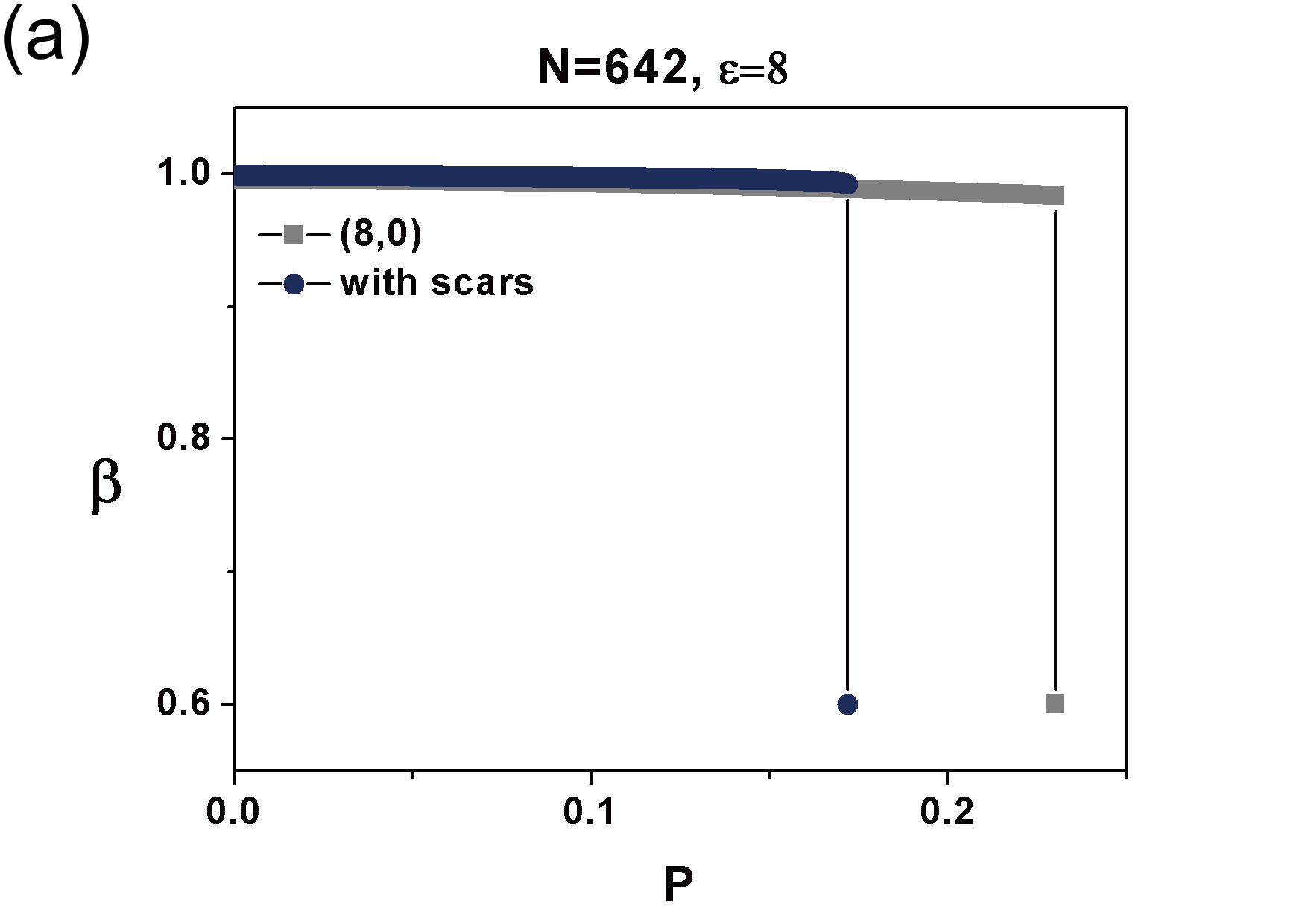} 
\hspace{1cm}
\includegraphics[width=3in]{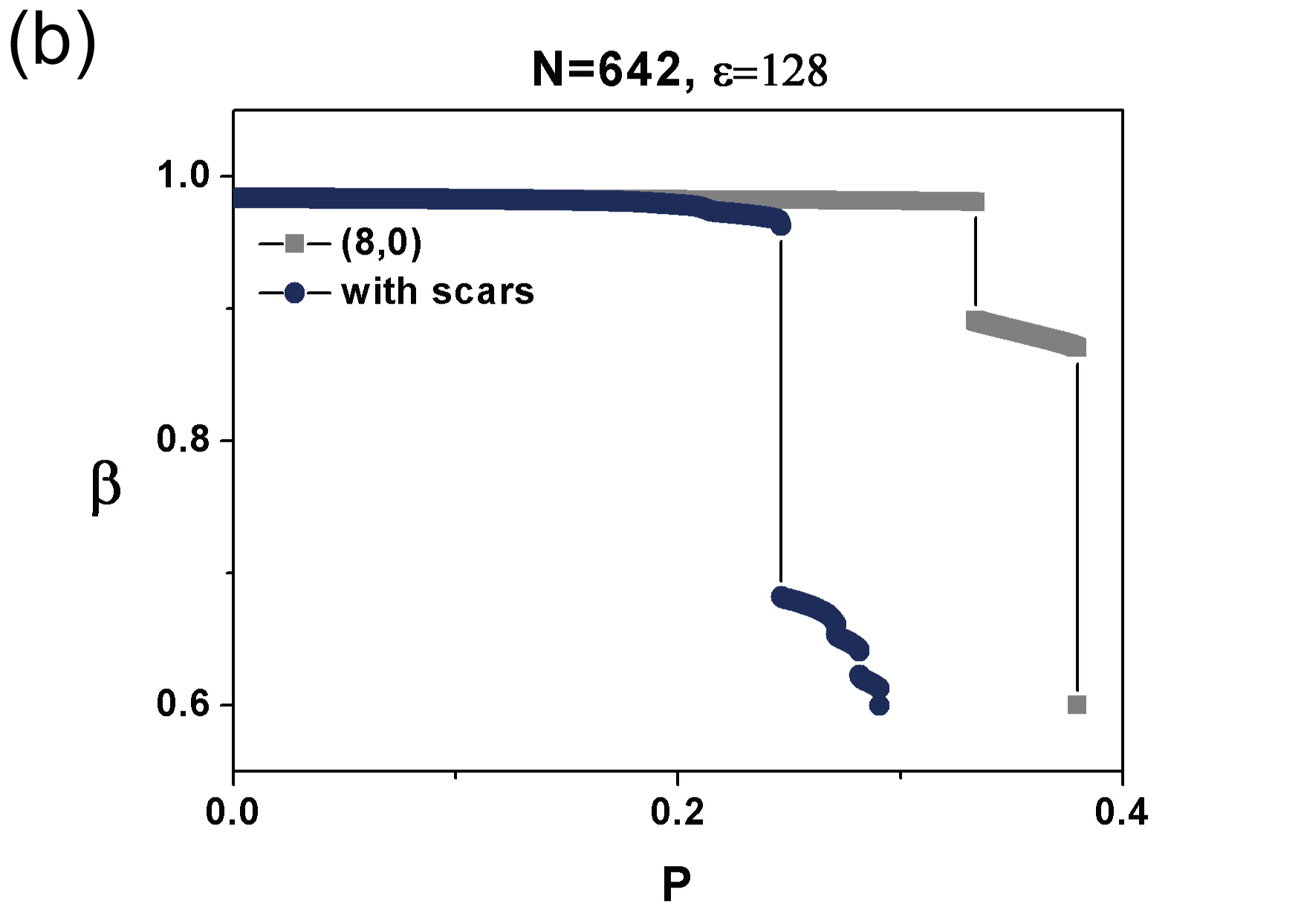} 
\caption{(Color online) Evolution of the parameter $\beta$ under the external pressure. (a) For small values of $\varepsilon$, shells collapse directly to $\beta_{min}$.  (b) For large $\varepsilon$, shells collapse
in stages. In the latter case, the critical pressure is defined as the point of the first collapse.} 
\label{evolution}
\end{figure*}

\begin{figure*}
\centering 
\includegraphics[width=3in]{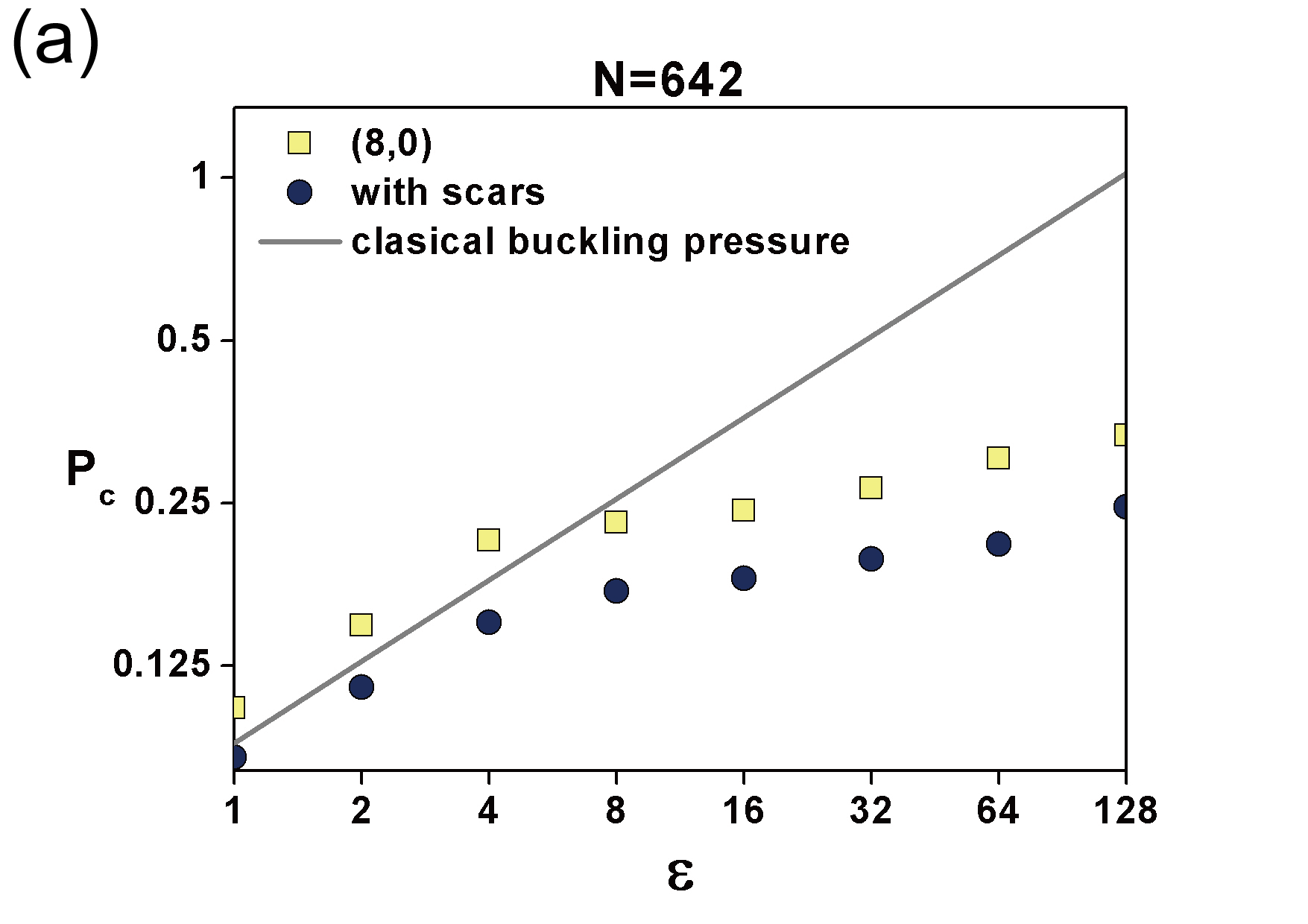} 
\hspace{1cm}
\includegraphics[width=3in]{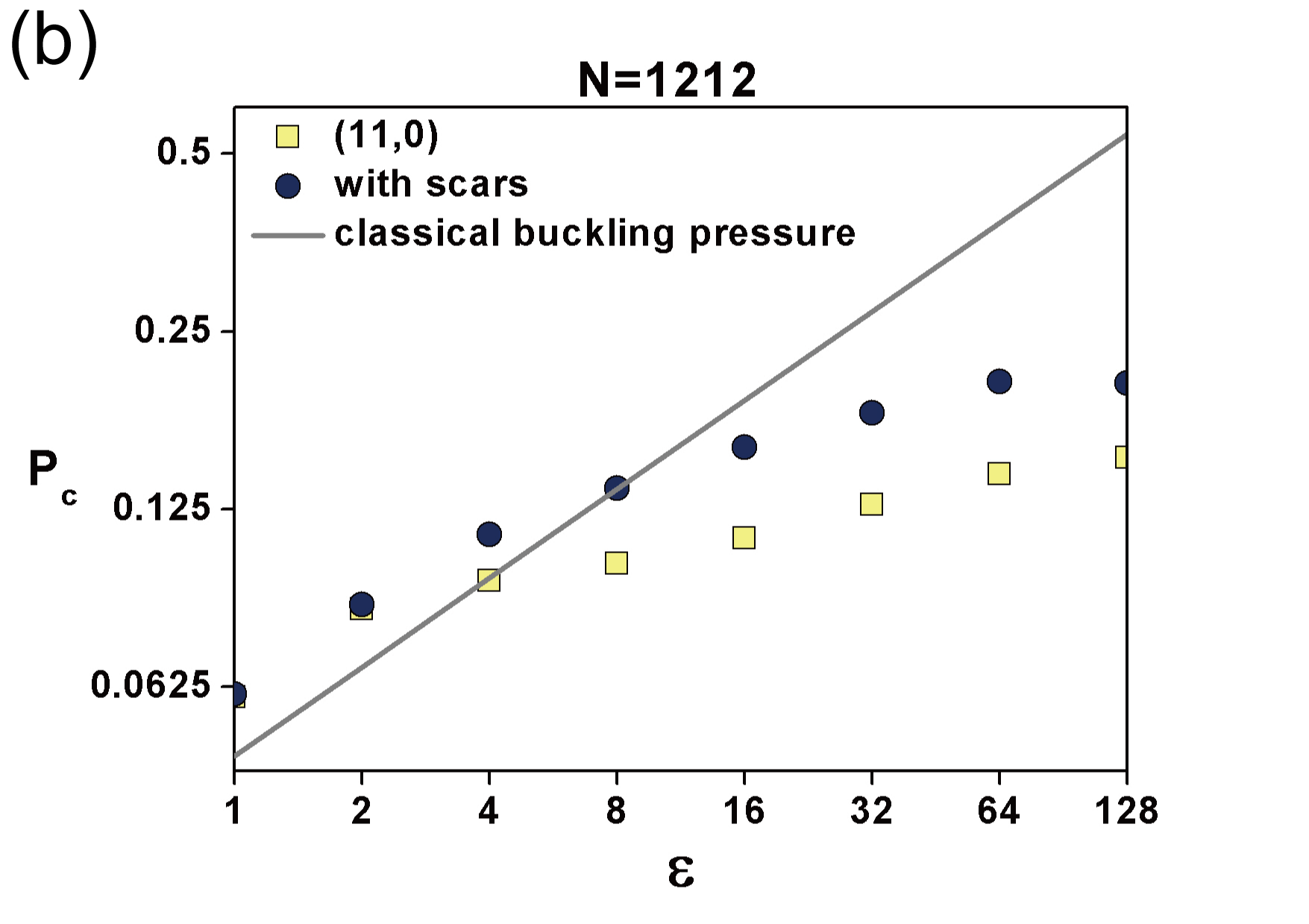} 
\caption{(Color online) Critical pressure $P_c$ for the $N_{1}=642$ shells (a) and the $N_{2}=1212$ shells (b). The grey lines indicate the buckling pressure of ideal
spherical shells as predicted by continuum elastic theory.} 
\label{critical pressure}
\end{figure*}

\begin{figure*}
\centering 
\includegraphics[width=7in]{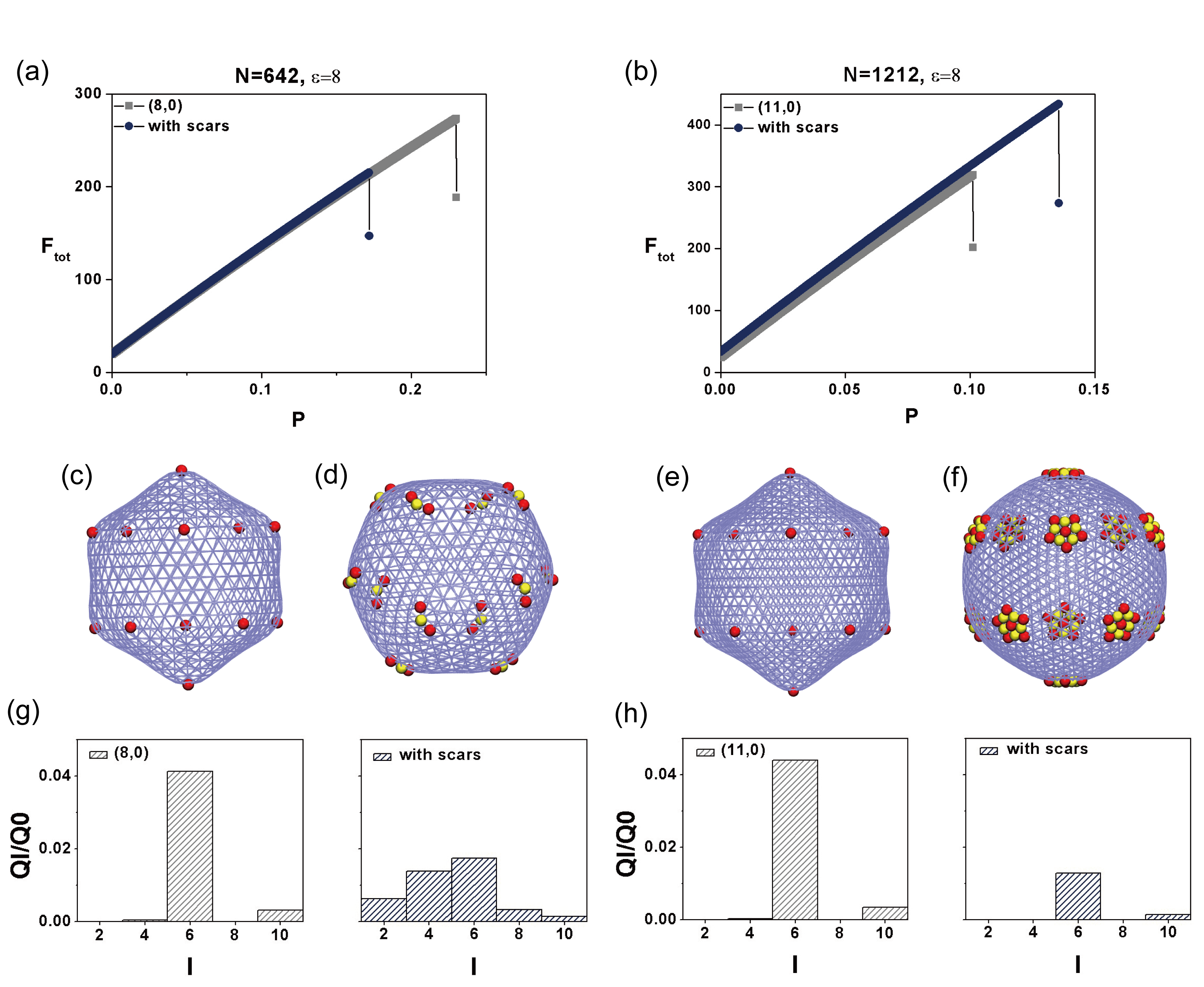} 
\caption{(Color online) Plots of the evolution of the system energy $F_{tot}$ for $N_{1}=642$ shells with $\varepsilon=8$ (a) and $N_{2}=1212$ shells with $\varepsilon=8$ (b). (c)-(f) Corresponding 
configurations for shells before collapse.  (g)-(h) Histograms of Ql/Q0 for the configurations.} 
\label{configurations}
\end{figure*}

In the presence of an external hydrostatic pressure $P$, an additional term, $PV$, has to be added to the elastic energy functional, such that $F_{tot} = F_{el} + PV$.  To quantify
the deformation of shape, we first introduce a dimensionless parameter $\beta = \left(36 \pi \right) ^{\frac{1}{6}}  V^{\frac{1}{3}}/A^{\frac{1}{2}}$, where $V$ and $A$ 
are volume and surface area, respectively. $\beta$ provides a quantitative measure of the extent of convexity of the shell, with $\beta=1$ for a sphere and $\beta \approx 0.969$ for an icosahedron. 
$\beta$ decreases as the shell deflates under pressure. 
In our simulations, we set a lower limit $\beta_{min}=0.6$ to avoid unphysical self-intersections
when the system is close to full collapse. We find that for small values of $\varepsilon$, shells collapse discontinuously to $\beta_{min}$, as shown in
Fig.~\ref{evolution}(a). The discontinuity defines a critical pressure $P_c$ at which the shell can no longer retain its shape and collapses. For larger values of $\varepsilon$ shells collapse in stages,
as shown in Fig.~\ref{evolution}(b). In this case we define $P_c$ as the pressure at the first collapse. For even larger values of $\varepsilon$ the collapse tends to become
continuous and a definition of the critical pressure becomes more ambiguous. Here we only consider values of $\varepsilon$ for which $P_c$ can be unambiguously defined. In the $N_{1}=642$ ($(p,q)=(8,0)$) case, shells 
without scars always have a higher value of $P_c$, as shown in Fig.~\ref{critical pressure}(a). In contrast, shells with scars appear to always have a higher critical 
pressure in the $N_{2}=1212$ ($(p,q)=(11,0)$) case, as shown in Fig.~\ref{critical pressure}(b). In both cases, in the regime of small $\varepsilon$, where the mean ``asphericity" does not depart significantly from zero in the absence of external pressure, the critical pressures are 
close to the buckling pressures $p \equiv 4 \sqrt{\kappa Y}/R^{2}$ of an ideal spherical shell, as predicted by continuum elastic theory \cite{van2009wt,paulose2012fluctuating}.

\section{Discussion}
\label{sec:discussion}

\begin{figure*}
\centering
\includegraphics[width=6in]{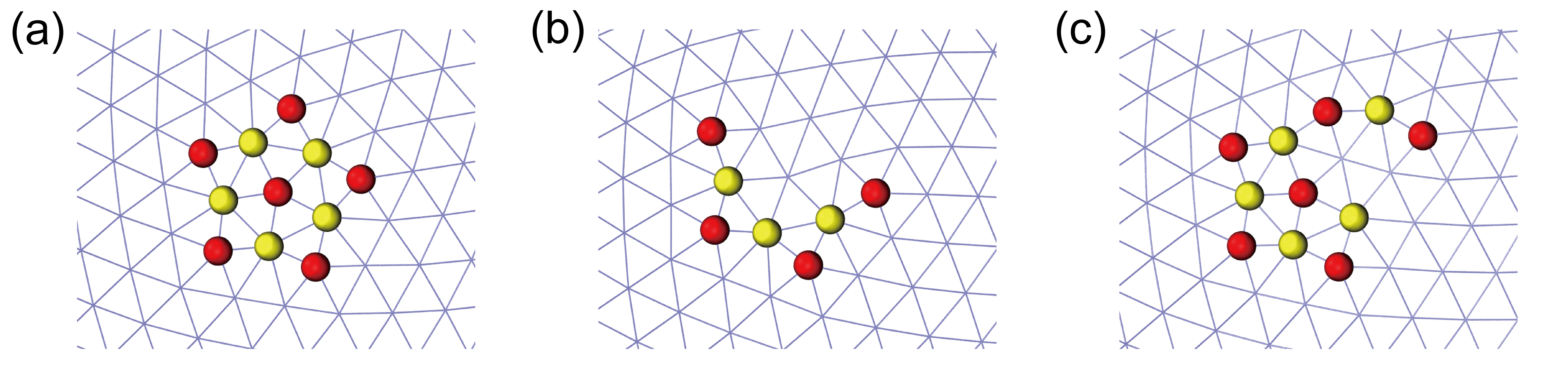}
\caption{(Color online) (a) One of the twelve star-like scars. (b)-(c) Two ways of breaking the icosahedral symmetry by flipping a bond. } 
\label{scar}
\end{figure*}

To gain a better understanding of the critical pressure, as an example we show in Fig.~\ref{configurations} the evolution of the system energy $F_{tot}$ with $\varepsilon=8$ along with the corresponding shell configurations before collapse.
The system without scars has a higher energy before collapse in the $N_{1}=642$ case, while the system with scars has a higher energy
before collapse in the $N_{2}=1212$ case. 
The configurations suggest that shells preserve their original symmetry, e.g., the symmetry in the absence of pressure, until collapse.
A more quantitative way to look at it is through spherical harmonic expansions \cite{lidmar2003virus, yong2013elastic} 
\begin{eqnarray}
R\left(\theta, \phi \right) &= &\sum_{i}  R_{i} \delta \left( \phi-\phi_{i} \right) \delta \left(\cos \theta - \cos \theta_{i} \right) \nonumber \\
                            &= &\sum_{l,m}Q_{lm}Y_{lm}\left(\theta, \phi \right),
\end{eqnarray}
where $\left(R_{i}, \theta_{i}, \phi_{i} \right)$ represents the spherical coordinates of vertex $i$ in the reference frame centered at the shell's center of mass.
From the coefficients $Q_{lm}$, one may form two rotationally invariant quantities \cite{steinhardt1983bond}
\begin{eqnarray}
Q_{l}=\left( \frac{4\pi}{2l+1} \sum_{m=-l}^{l} |Q_{lm} |^2 \right)^{1/2}
\end{eqnarray} 
and
\begin{eqnarray}
\hat{W}_{l} =\frac{\sum\limits_{m_{1},m_{2},m_{3} \atop m_{1}+m_{2}+m_{3}=0 } \left( \begin{array}{ccc} 
   l & l & l \\
   m_{1} & m_{2} & m_{3} \\
\end{array} \right) Q_{lm_{1}}Q_{lm_{2}}Q_{lm_{3}} }{\left( \sum\limits_{m} \left| Q_{lm} \right|^{2} \right)^{3/2}} ~. \nonumber \\
\end{eqnarray} 
The coefficients $ \left( \begin{array}{ccc} 
   l & l & l \\
   m_{1} & m_{2} & m_{3} \\
\end{array} \right)$ are Wigner $3j$ symbols \cite{landau1977quantum}. For a shape with the icosahedral symmetry, the $\lbrace Q_{l} \rbrace$ are nonzero only for $l$ = 0, 6, 10, ... \cite{steinhardt1983bond}.
The magnitudes of $\lbrace Q_{l} \rbrace$ change with shape. The $\lbrace \hat{W}_{l} \rbrace$ are normalized so that they are independent of the overall magnitude of the $\lbrace Q_{lm} \rbrace$ 
for a given $l$. The parameters $|\hat{W}_{l}|$  are a direct index of the symmetry of a shape (see Ref.~\cite{steinhardt1983bond} and references therein). In Fig.~\ref{configurations} we plot $Q_{l}/Q_{0}$ 
and in Table \ref{Wl} we compare $|\hat{W}_{l}|$  with the characteristic values for an icosahedral symmetry given in Ref.~\cite{steinhardt1983bond}. Except for small errors that are likely to be due to numerical
accuracy, the results agree well with previous findings. In the $N_{2}=1212$ case, both shells preserve icosahedral symmetry before collapse; with bonds of five-fold vertices compressed, 
as can be seen from Fig.~\ref{scar}(a), scars correspond to harder areas, thus resulting in a reduced effective radius and a higher critical pressure.

\begin{table}
\centering
\caption{$|\hat{W}_{l}|$ for icosahedron and shapes in Fig \ref{configurations}.}\label{tab:table}
\begin{tabular}{L{3cm}C{2.5cm}C{2.5cm}}
\hline 
\hline

                     &   $|\hat{W}_{6}|$  & $|\hat{W}_{10}|$\\
\hline
Icosahedron          &    0.169754      &   0.093967\\
(8,0)                &    0.169751      &   0.093951\\
N=642,with scars     &    0.117793      &   0.002383\\
(11,0)               &    0.169751      &   0.093964\\
N=1212,with scars    &    0.169751      &   0.093968\\
\hline
\end{tabular}
\label{Wl}
\end{table}

\begin{figure}
\centering
\includegraphics[width=3.5in]{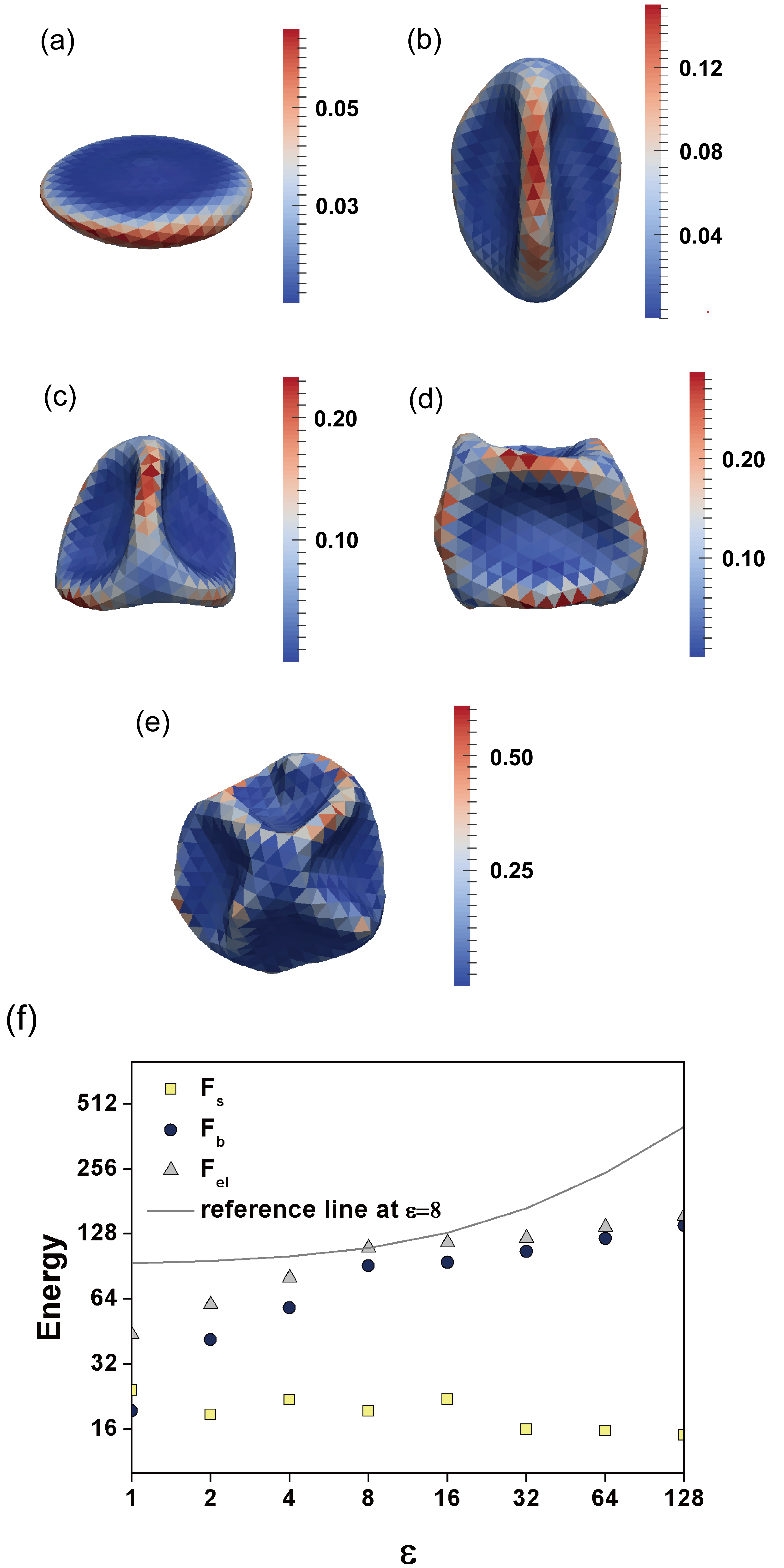}
\caption{(Color online) (a)-(e) Configurations of $(8,0)$ shells at $\beta=0.8$ during collapsing, with $\varepsilon=1, 2, 4, 8, 128$, respectively.
Color represents local bending energy. (f) Energies plotted vs $\varepsilon$. The reference line at $\varepsilon=8$ shows the elastic energy keeping the configuration at $\varepsilon=8$ frozen while varying $\varepsilon$.  } 
\label{collapse}
\end{figure}

Therefore, we argue that two mechanisms affect the critical pressure: i) the preservation of icosahedral symmetry and ii) the area occupied by scars. As a further check, 
we break the icosahedral symmetry of configurations with scars by flipping a bond of one of the twelve star-like scars, in two ways shown in Fig. \ref{scar}(b) and (c), 
and find the critical pressures are lowered significantly in both situations. We test shells with a $\left(p, q\right)$ structure and with the total number of vertices in
the $400$ to $2000$ range (see Appendix for details) and find results that align well with expectations.

During the collapse, we find emergence of configurations with a broken icosahedral symmetry. Fig.~\ref{collapse} shows configurations at $\beta=0.8$. 
With $\varepsilon$ increasing, the shell at first appears to retain a regular shape, e.g., a disk, a three-fold deflated American football, a concave tetrahedron or a squashed 
cube. The appearance of these shapes is not surprising since the icosahedral symmetry contains six 5-fold axes, ten 3-fold axes and fifteen 2-fold axes. In addition, an icosahedron
can be decomposed into three orthogonal rectangular planes \cite{steinhardt1983bond}. As $\varepsilon$ increases further, irregular shapes with more ridges start to appear. 
These account for most of the bending energy but help to lower the overall stretching energy. A Landau-like theory of phase transitions has been constructed
for the emergence of a cubic-like shape during the collapse of an icosahedral shell \cite{yong2013elastic}. In the energy plot Fig.~\ref{collapse}(f), the reference line at $\varepsilon=8$ lies above the
elastic energy at other $\varepsilon$ values. Except for small fluctuations, we find the situation is similar for references lines at other values of $\varepsilon$, 
which indicates the configuration at a certain $\varepsilon$ is favored compared with other configurations.

\section{Summary and conclusions}
\label{sec:summary}

\begin{table}
\centering
\caption{Symmetries of the corresponding scarred shells.}\label{tab:table}
\begin{tabular}{L{1.0cm}C{1.0cm}C{1.5cm}|L{1.0cm}C{1.0cm}C{1.5cm}} 
\hline \hline
  $N$     & $\left(p,q\right)$ & point group & $N$  & $\left(p,q\right)$ & point group\\
\hline
432 & $\left(6, 1 \right)$ & $D_{3}$    & 1242 & $\left(10, 2 \right)$ & $C_{1}$ \\ 
482 & $\left(4, 4 \right)$ & $C_{2}$    & 1272 & $\left(7, 6 \right)$ & $T$  \\ 
492 & $\left(5, 3 \right)$ & $C_{2}$    & 1292 & $\left(8, 5 \right)$ &  $=$        \\ 
492 & $\left(7, 0 \right)$ & $C_{2}$    & 1332 & $\left(9, 4 \right)$ &  $I$        \\ 
522 & $\left(6, 2 \right)$ & $C_{2}$    & 1332 & $\left(11, 1 \right)$ & $I$        \\ 
572 & $\left(7, 1 \right)$ & $D_{3}$    & 1392 & $\left(10, 3 \right)$ & $C_{1}$  \\ 
612 & $\left(5, 4 \right)$ & $C_{2}$    & 1442 & $\left(12, 0 \right)$ & $C_{1}$  \\ 
632 & $\left(6, 3 \right)$ & $D_{3}$    & 1472 & $\left(11, 2 \right)$ & $C_{2}$  \\ 
642 & $\left(8, 0 \right)$ & $C_{2}$    & 1472 & $\left(7, 7 \right)$ &  $C_{2}$  \\ 
672 & $\left(7, 2 \right)$ & $C_{1}$    & 1482 & $\left(8, 6 \right)$ &  $T$  \\ 
732 & $\left(8, 1 \right)$ & $T$        & 1512 & $\left(9, 5 \right)$ &  $I$         \\ 
752 & $\left(5, 5 \right)$ & $C_{2}$    & 1562 & $\left(10, 4 \right)$ & $=$         \\ 
762 & $\left(6, 4 \right)$ & $C_{2}$    & 1572 & $\left(12, 1 \right)$ & $I$         \\ 
792 & $\left(7, 3 \right)$ & $T$        & 1632 & $\left(11, 3 \right)$ & $I$         \\ 
812 & $\left(9, 0 \right)$ & $C_{2}$    & 1692 & $\left(8, 7 \right)$ &  $I$        \\ 
842 & $\left(8, 2 \right)$ & $C_{1}$    & 1692 & $\left(13, 0 \right)$ & $I$         \\ 
912 & $\left(6, 5 \right)$ & $C_{1}$    & 1712 & $\left(9, 6 \right)$ &  $=$        \\ 
912 & $\left(9, 1 \right)$ & $C_{1}$    & 1722 & $\left(12, 2 \right)$ & $=$         \\ 
932 & $\left(7, 4 \right)$ & $C_{1}$    & 1752 & $\left(10, 5 \right)$ & $I$         \\ 
972 & $\left(8, 3 \right)$ & $T_{h}$    & 1812 & $\left(11, 4 \right)$ & $=$         \\ 
1002 & $\left(10, 0 \right)$ & $T$      & 1832 & $\left(13, 1 \right)$ & $I$          \\ 
1032 & $\left(9, 2 \right)$ & $C_{1}$   & 1892 & $\left(12, 3 \right)$ & $=$         \\ 
1082 & $\left(6, 6 \right)$ & $D_{5}$   & 1922 & $\left(8, 8 \right)$ &  $=$       \\ 
1092 & $\left(7, 5 \right)$ & $T$   & 1932 & $\left(9, 7 \right)$ &  $I$        \\ 
1112 & $\left(10, 1 \right)$ & $C_{1}$  & 1962 & $\left(10, 6 \right)$ & $=$         \\ 
1122 & $\left(8, 4 \right)$ & $C_{1}$   & 1962 & $\left(14, 0 \right)$ & $=$      \\ 
1172 & $\left(9, 3 \right)$ & $C_{1}$   & 1992 & $\left(13, 2 \right)$ & $=$         \\ 
1212 & $\left(11, 0 \right)$ & $I$      &     &                       &          \\ 
\hline \hline 
\end{tabular}
\label{symmetry}
\end{table}

We used numerical simulations to show that the topological scars present in the crystalline lattice of thin shell with spherical topology affect shell's 
ability to sustain external hydrostatic pressure. Furthermore, we demonstrated that the distribution of the scars within the lattice, i.e., their symmetry,  
non-trivially affects shells' resistance to pressure. We find that the critical pressure at which shells collapse is lowered when the scar distribution breaks the icosahedral
symmetry and raised when symmetry is preserved. We find that the isotropic pressure will not alter the symmetry of shells before collapse. The emergence of shapes with
a broken icosahedral symmetry during collapse for shells starting with icosahedral symmetry is a function of $\varepsilon/\tilde{\kappa}$ (FvK number $\gamma$ with fixed $R$),
with more ridges that concentrate bending energy present to lower the stretching energy for high $\varepsilon/\tilde{\kappa}$.

This is yet another example of a problem in which the frustration caused by the incompatibility between order and underlying geometry causes very non-trivial behavior, with a
rich variety of shape patterns with no flat space analogues. In this study we have concentrated on a somewhat simpler case of frozen defects and only examine response of the
geometry to the external perturbation. It would be interesting, yet substantially more complicated, to explore how the defect distribution evolves as a shell deforms.

\acknowledgements  
DW thanks M. Xiao for constructive discussions. Numerical simulations were performed using the Syracuse University campus condor pool. This research was supported by the Soft Matter Program of Syracuse University.

\appendix
\section{Symmetries of various shell configurations}

Table \ref{symmetry} shows the $(p,q)$ pairs of shells without scars, with the number of vertices $N$ in the range from $400$ to $2000$ and the point groups the corresponding scarred configurations belong to \cite{wales2006structure}, with the connectivity matrix taken from the Thomson problem database \cite{thomson_applet}. $``="$ indicates the configuration with 
scars is the same as the configuration without scars, i.e., only twelve disclinations present. In all cases except the $(8,3)$ case, we find that within a clear definition, the critical pressure
at which shells collapse is lowered for scars that break icosahedral symmetry and raised for scars that preserve icosahedral symmetry. In the $(8,3)$ case, 
shells with scars have a higher critical pressure for certain $\varepsilon$ range, we think this is partly due to $T_{h}$ being a high order symmetry and 
partly due to it having a relatively large scar area.

\bibliographystyle{apsrev4-1}
\bibliography{references}

\end{document}